\documentclass[twocolumn,prl,superscriptaddress,showpacs]{revtex4}

\usepackage{graphicx}
\usepackage{amssymb}
\usepackage{amsfonts}

\DeclareGraphicsExtensions{.eps}

\begin{document}

\title{Parametric polariton amplification in semiconductor microcavities}

\date{\today}

\author{G. Messin}
\affiliation{Laboratoire Kastler Brossel, Universit\'{e} Paris 6, Ecole
Normale Sup\'{e}rieure et CNRS,\\
UPMC Case 74, 4 place Jussieu, 75252 Paris Cedex 05, France}

\author{J.Ph. Karr}
\affiliation{Laboratoire Kastler Brossel, Universit\'{e} Paris 6, Ecole
Normale Sup\'{e}rieure et CNRS,\\
UPMC Case 74, 4 place Jussieu, 75252 Parisk Cedex 05, France}

\author{A. Baas}
\affiliation{Laboratoire Kastler Brossel, Universit\'{e} Paris 6, Ecole
Normale Sup\'{e}rieure et CNRS,\\
UPMC Case 74, 4 place Jussieu, 75252 Paris Cedex 05, France}

\author{G. Khitrova}
\affiliation{Optical Sciences Center, University of Arizona, Tucson, AZ 85721-0094,USA.}

\author{R. Houdr\'{e}}
\affiliation{Institut de Micro et Opto\'{e}lectronique, Ecole Polytechnique F\'{e}d\'{e}rale de
Lausanne, Lausanne, CH 1015 Switzerland}

\author{R.P. Stanley}
\affiliation{Institut de Micro et Opto\'{e}lectronique, Ecole Polytechnique F\'{e}d\'{e}rale de
Lausanne, Lausanne, CH 1015 Switzerland}

\author{U. Oesterle}
\affiliation{Institut de Micro et Opto\'{e}lectronique, Ecole Polytechnique F\'{e}d\'{e}rale de
Lausanne, Lausanne, CH 1015 Switzerland}

\author{E. Giacobino}
\affiliation{Laboratoire Kastler Brossel, Universit\'{e} Paris 6, Ecole
Normale Sup\'{e}rieure et CNRS,\\
UPMC Case 74, 4 place Jussieu, 75252 Paris Cedex 05, France}

\begin{abstract}
We present novel experimental results demonstrating the coherence properties of the nonlinear
emission from semiconductor microcavities in the strong coupling regime, recently interpreted by
parametric polariton four-wave mixing. We use a geometry corresponding to degenerate four-wave
mixing. In addition to the predicted threshold dependence of the emission on the pump power and
spectral blue shift, we observe a phase dependence of the amplification which is a signature of a
coherent polariton wave mixing process.
\end{abstract}


\maketitle

In high finesse semiconductor microcavities with embedded quantum wells\cite{cargese}, the photon
and exciton confinement and the large excitonic oscillator strength make it possible to reach the
strong coupling regime or normal mode coupling. As a result, the degeneracy at resonance between
the exciton mode and the photon mode is lifted and the so-called vacuum Rabi splitting takes
place\cite{weisbuch}. The resulting two-dimensional eigenstates, called cavity polaritons, are
mixed photon-exciton states that have a number of attractive features. They are considered as
potential candidates for Bose-Einstein condensation because of the occurrence of strong
polariton-polariton scattering into the small $k_{\parallel }$ states. The purpose of this paper is
to report on novel properties of cavity polaritons in the nonlinear regime, that are a direct proof
of the coherent nature of the polariton-polariton interaction.

The emission intensity of these systems has been shown to undergo a giant amplification when the
intensity of the driving field is increased above some threshold, both under non
resonant\cite{cao,lesidang,bloch} and resonant laser
excitation\cite{savvidis,huang,houdre,stevenson}. Several mechanisms were proposed to explain this
behavior, including stimulated exciton phonon scattering and stimulated exciton exciton scattering,
raising a lot of controversy\cite{immamoglu,tassone,kira}. In particular, experiments using a
resonant pump at a specific angle and measuring the amplified emission or the gain on a probe laser
at normal incidence \cite{savvidis,houdre,stevenson}, are in good agreement with a recent
theoretical model based upon coherent polariton four-wave mixing \cite{ciuti}. A strong
amplification is observed when the angle is chosen in order to ensure energy and in-plane momentum
conservation for the process where two pump polaritons (with momentum $k_{P\parallel }$) are
converted into a signal polariton ($k_{\parallel }=0$) and an idler polariton ($k_{\parallel
}=2k_{P\parallel }$).

The experiments involving resonant excitation have been performed in a non-degenerate
configuration, in which the signal, idler and pump beams have different energies and momenta. An
alternative configuration ensures the double energy and momentum resonance~: it is the one where
$k_{P\parallel }=0 $, with $k_{signal\parallel }=k_{idler\parallel }=0,$ the energy of the pump
laser being resonant with the polariton energy. In this paper, we present experimental results
obtained in this configuration. The amplified emission, which now contains both signal and idler,
is observed in the direction of the reflected pump beam. The ``degenerate'' configuration allows us
to highlight new features of the polariton parametric interaction. To analyze the characteristics
of the emission in detail, a highly sensitive homodyne detection is used. As in the non-degenerate
case, a marked threshold is observed when the pump intensity is increased and the associated shift
is found to be in agreement with ref. \cite{ciuti}. Degenerate four-wave mixing has been studied by
several authors \cite {kuwata,quochi} in semiconductor microcavities, however in different regimes
where the giant amplification on the lower polariton branch did not occur.

Moreover, it is well known that in this geometry optical parametric amplification is phase
sensitive\cite{yariv}. Using a single mode cw laser as a pump, we have been able to investigate for
the first time the phase dependence of the giant emission. If it was due to stimulation by the
lower polariton occupation number\cite{stevenson}, a phase insensitive amplification would be
observed, as in a laser-like amplifier\cite{optamp}. We have observed that the amplification
depends strongly on the phase of the pump laser. This result brings in a crucial argument in favor
of the interpretation of the amplified emission as originating primarily from coherent polariton
four-wave mixing in the case of resonant pumping. Let us note that the case of non resonant
pumping\cite{bloch} where there is no phase reference in the system corresponds to a very different
problem which requires a specific approach.

We use high finesse GaAs/AlAs microcavity samples containing either one or two InGaAs quantum wells
with low indium content. The samples are described respectively in refs.\cite{houdre} (sample 1)
and \cite{nmc} (sample 2). These high quality samples exhibit very narrow polariton linewidth, on
the order of 100-200~$\mu $eV (half-width at half-maximum) at 4~K. The microcavities are wedged,
allowing one to change the cavity thickness by moving the laser excitation spot on the samples. The
beam of a single mode cw Ti:sapphire laser with a linewidth of the order of 1~MHz is focused onto
the sample at normal incidence. The laser light is circularly polarized. The laser spot has a
diameter of 80$~\mu $m.

First the positions of the two polariton branches are determined by studying the reflection of the
driving laser at very low intensity ($I$~$\lesssim $~1~mW, corresponding to 10W/cm$^{2}$). For
sample 2, this yields the diagram shown in Fig.~\ref{fig2}, where the energies of the reflectivity
minima have been plotted as a function of their positions on the sample. A similar diagram has been
obtained for sample 1, with a smaller vacuum Rabi splitting.

\begin{figure}[t]
\includegraphics[scale=0.45]{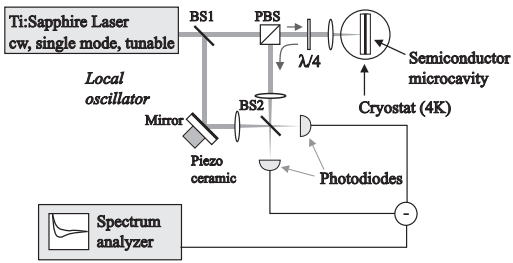}
\caption{Experimental set-up. It can be used in two different configurations: (a) Without
beamsplitter (BS1): The polarizing beamsplitter (PBS) and the quarter wave plate ($\lambda /4$)
form an optical circulator that direct all the circularly polarized light coming from the
microcavity towards the detection system. The 50/50 beamsplitter (BS) splits the beam into two
equal parts which fall on two photodetectors. The \textit{sum} of the two photocurrents gives the
beat signal between the emission and the laser. The \textit{difference} between the photodiode
currents yields the shot noise. (b) With beamsplitter (BS1): an independent local oscillator is
split off by the beamsplitter (BS1) and mixed with the beam emerging from the microcavity on (BS2).
In this case, the beat signal between the emitted light and the local oscillator is obtained by
taking the \textit{difference} between the two photocurrents. The shot noise background is measured
by blocking the beam emerging from the cavity\cite{note}.} \label{fig1}
\end{figure}

The light emitted by the microcavity is detected in the direction normal to the sample by means of
a homodyne detection system\cite{yamamoto,yuen}. For this purpose the emitted light is mixed with a
local oscillator on a photodetector (Fig. \ref{fig1}). The emitted light of interest copropagates
with the laser beam and has the same very small divergence angle ($\simeq $0.4${{}^{\circ}}$),
ensuring the $k_{\parallel }$ conservation. An additional laser beam, the local oscillator, is
mixed with the pump and signal beam on a beamsplitter. The beams coming from the beamsplitter are
focused on two photodetectors. The frequency spectrum of the photocurrents is analyzed with an RF
spectrum analyzer. The reflected laser and the light scattered at the laser frequency yield a large
peak at zero frequency which is filtered out. The signal given by the spectrum analyzer can be
shown\cite{yuen} to be proportional to the beat signal between the local oscillator and the light
emitted by the sample, i.e. luminescence. For an incoherent emission, the spectrum analyzer
measures a broad signal, that appears on top of the background quantum noise (shot noise).

\begin{figure}[b]
\includegraphics[scale=0.45]{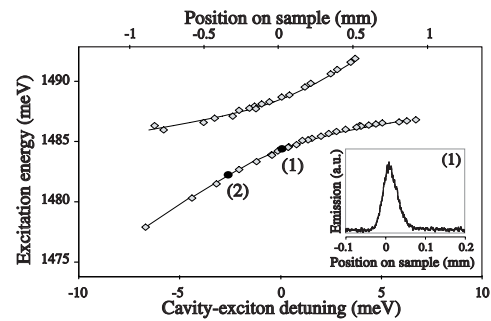}
\caption{Energy diagram of the polariton branches as a function of position on sample 2. Inset:
emission resonance obtained when the laser position is scanned on sample 2 around point (1) with
fixed pump laser wavelength and $I_{pump}=0.7mW$. Here and in the following figures, the analysis
frequency of the spectrum analyzer is fixed at 7 MHz.} \label{fig2}
\end{figure}

In a first set of experiments, the local oscillator is the laser beam reflected by the microcavity.
The phase difference between the detected signal and the local oscillator is fixed, since the
emitted light and the reflected laser follow exactly the same path. This scheme is well adapted to
investigate the amplification threshold and the line shift. In a second set of experiments, we have
used an independent local oscillator beam, split off from the pump laser. One of the mirrors on the
path of the local oscillator is mounted on a piezoceramic, which allows us to vary its phase
relative to the signal beam. This provides a tool to explore all the quadrature components for the
emitted light. Let us stress that without homodyne detection the emitted light would be extremely
difficult to distinguish from the scattered and reflected laser light.

We first describe the results concerning the emission threshold and the line shift, using the first
experimental scheme. In order to study the emission resonances we had to use a specific method. The
possible frequency scan of the detection system on each side of the laser line is of the order of a
few tens of MHz, that is in the 100 neV range, much smaller than a typical photoluminescence
spectrum. To investigate the emission, the frequency of the spectrum analyzer is kept fixed, while
either the laser frequency or the position on the sample is scanned. Therefore, the detected
polaritons have an energy extremely close to the ones excited by the pump laser. The emission
lineshape obtained with the homodyne detection can be studied in the vicinity of the polariton
resonance branches either by scanning the laser energy (``vertical'' scan in Fig.~\ref{fig2}) for
fixed positions on the sample or by scanning the position of the laser on the sample
(``horizontal'' scan in Fig.~\ref{fig2}) at fixed laser frequencies. Since it is difficult to scan
a single mode laser continuously over ranges that are of the order of the polariton linewidth, we
chose the second procedure. The position scan amounts to varying the detuning between the polariton
and the laser at a fixed laser frequency. An example of such a signal is shown in the inset of
Fig.~\ref{fig2} for very low laser intensity.

\begin{figure}[t]
\includegraphics[scale=0.65]{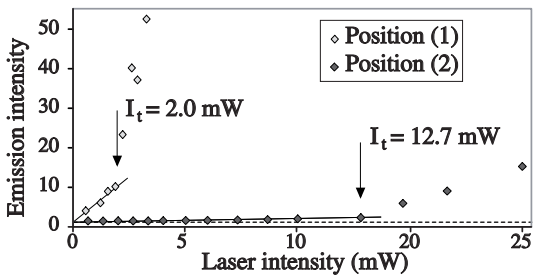}
\caption{Emission intensity as a function of pump laser intensity for positions (1) and (2) on the
lower polariton branch shown in Fig.~\ref{fig2}. $I_{t}$ is the threshold driving laser intensity.
The laser frequency is fixed and the position on the sample is adjusted to maximized the signal.}
\label{fig3}
\end{figure}

At very low driving intensity, the emission exhibits maxima for the same laser frequencies and
positions on the sample as the reflectivity minima\cite {stanley}. We have studied the height and
position of the emission resonance as a function of the driving laser intensity. Figure \ref{fig3}
shows the maximum intensity versus the pump laser intensity for two different positions on the
lower polariton branch of sample 2 (points (1) and (2) in Fig~\ref{fig2}), corresponding to two
different laser frequencies. The emission intensity exhibits a marked threshold. Below threshold
the emitted intensity has linear dependence on the intensity of the driving laser, as is expected
from thermal luminescence. Above threshold, the emitted intensity increases very fast with the
driving intensity. The threshold takes place at very low excitation intensities. It is as low as
2~mW of incident laser power over a 80~$\mu $m spot (20~W/cm$^{2}$). The threshold is lowest for
values of the cavity-exciton detuning close to zero. As shown in Fig.~\ref{fig3}, it is higher for
point (2) where the exciton content of the polariton is lower. All the considered pump intensities
correspond to exciton densities that are below the polariton bleaching density. This was verified
by checking the presence of vacuum Rabi splitting in reflectivity for these intensities. No
nonlinear emission was observed on the upper polariton branch in the same range of laser powers.

We have also studied the shift of the emission curve as a function of the laser intensity. We find
that the position on the sample of the maximum of the emission curve varies linearly with the
intensity of the driving laser below threshold. The curves shift to the left of the unperturbed
lower branch resonance in Fig.~\ref{fig2}, which is equivalent to a blue shift in energy. The sign
of the shift is consistent with a non-linearity coming from polariton-polariton scattering, which
is repulsive. The measured value of the shift at threshold for point (1) in the level diagram is
0.24~meV, in good agreement with the value of about twice the polariton linewidth $2\gamma
=0.29~$meV predicted by ref.\cite{ciuti}. The linewidth (HWHM) $\gamma $ was measured independently
using the reflectivity resonances. These results are consistent with those of previous experiments,
obtained in the non-degenerate geometry.

In the case of degenerate operation, with the signal and idler having the same energies and
wavevectors, the parametric amplifier is expected to be phase-dependent. Using homodyne detection,
we are able to explore this property, in contrast to other authors. As in the experiment performed
in ref.\cite {houdre}, thermally excited polaritons constitute the probe that seeds the parametric
process and that is amplified through interaction with the pumped polaritons. Having no average
phase, the former can be considered as a superposition of random polariton fields with equal mean
amplitudes and phases spread over 2$\pi$. For each value of the local oscillator phase, the
emission having a specific phase is singled out by the homodyne detection. Scanning the local
oscillator phase allows us to identify whether some quadrature components are amplified in a
preferential way.

\begin{figure}[t]
\includegraphics[scale=0.75]{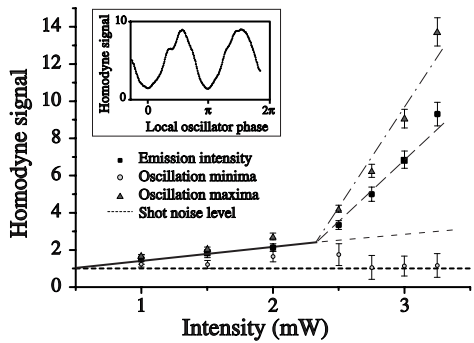}
\caption{Amplitude of the minima and maxima of the oscillations as a function of the driving laser
intensity; the driving laser frequency is fixed and equal to the unshifted (i.e. measured at very
low intensity) lower polariton frequency at zero cavity-exciton detuning; the position on sample 1
is adjusted to maximize the emission intensity; dash-dotted and dashed lines: guides to the eye
respectively for the oscillation maxima and the emission intensity above threshold; solid line:
thermal emission intensity below threshold, obtained as in Fig~\ref{fig3}; dotted line:
extrapolated thermal emission. Inset: homodyne detection signal (arbitrary units) when the phase of
the local oscillator is varied, for a driving laser power of 3mW. The oscillation period is $\pi $
rather than $2\pi $ because the probe is constituted by an ensemble of random fields in which
phases $\phi $ and $\phi +\pi $ are equivalent.} \label{fig4}
\end{figure}

Well below threshold, as expected, all quadrature components are equivalent. Starting a little
below threshold, one observes an oscillation of the homodyne signal as a function of the local
oscillator phase. A typical recording is shown in Fig.~\ref{fig4} (inset), at zero cavity-exciton
detuning for sample 1. This clearly demonstrates the phase dependence of the polariton field
typical of degenerate parametric amplification\cite{yariv}. The polariton-polariton interaction
that is at the origin of the observed amplification is a coherent one, which can be identified with
polariton four-wave mixing. Indeed, the Hamiltonian annihilating two pump polaritons $a_p^{}$ and
creating two identical signal photons $a_s^{}$ is of the form $V a_p^{}\, a_p^{}\, a_s^{\dag}\,
a_s^{\dag}$, with $V$ the interaction energy. Then the semi-classical evolution equation for
$\langle a_s \rangle$ contains a gain term proportional to ${\langle a_p^2 \rangle} {\langle a_s
\rangle}^{*}$, which implies a well defined phase relationship between the pump field and the
amplified polariton quadrature. An amplification originating from population effects, as in a
laser, would be phase independent.

The amplitude of the oscillation grows rapidly with the pump field above threshold. In
Fig.~\ref{fig4}, we show the maxima and minima of the oscillations as a function of the driving
laser power for sample 1, obtained at fixed laser frequency and by shifting the sample in order to
compensate for the energy shift (the threshold is different from the one obtained with sample 2 for
a similar position, but the overall behaviors are very much the same for the two samples). The
level of thermal luminescence in the absence of nonlinear process, extrapolated from its value
below threshold is also shown. It can be seen that the oscillation minimum goes well below the
level of thermal luminescence, indicating a strong de-amplification of thermal polaritons. This
effect is also a specific feature of parametric interaction. If de-amplification is efficient
enough, squeezing of the emission below the shot noise level could be achieved. De-amplification
down to the shot noise level, i.e. to the vacuum noise is obtained here. The quadrature component
of the emitted field along the mean field of the pump laser (called emission intensity in
Fig.~\ref{fig3} and~\ref{fig4}), that is measured using the first experimental scheme, undergoes
intermediate amplification.

In conclusion we have presented new results on parametric amplification in semiconductor
microcavities. Using homodyne detection in a degenerate configuration, we have been able to show
evidence for specific features of polariton four-wave mixing in the microcavity. The amplification
threshold and the shift of the emission line observed in other experiments have been obtained in a
configuration very different from that of other authors. We have shown for the first time the phase
dependence of the amplification process in the degenerate configuration and the occurrence of
de-amplification, demonstrating the coherent origin of the process. Since in such nonlinear
interactions amplification and squeezing originate from the same physical effect, this opens the
way to the possibility of squeezing in semiconductor microcavity devices \cite{walls,messin}.
Further investigations are necessary to assess the feasible level of squeezing.

Acknowledgments. This work was performed with the support the European Commission (TMR ERMFMRX CT
96 00066) and of French DGA-DRET and CNRS (Ultimatech and Telecommunications). G. K. acknowledges
support from NSF AMOP.


\vspace{0.5 cm}

\begin{thebibliography}{99}

\bibitem{cargese}
H. Benisty \textit{et al.}, \textit{Confined Photon Systems} (Springer 1999)

\bibitem{weisbuch}
C. Weisbuch \textit{et al.}, Phys. Rev. Lett. \textbf{69} 3314 (1992); R. Houdr\'{e} \textit{et
al.}, Phys. Rev. Lett. \textbf{73} 2043 (1994)

\bibitem{cao}
H.\ Cao \textit{et al.}, Phys. Rev. A \textbf{55} 4632 (1997)

\bibitem{lesidang}
Le Si Dang \textit{et al.}, Phys. Rev. Lett. \textbf{81} 3920 (1998); F. Boeuf \textit{et al.},
Phys. Rev. B \textbf{62} R2279 (2000)

\bibitem{bloch}
P. Senellart and J. Bloch, Phys. Rev. Lett. \textbf{82} 1233 (1999)

\bibitem{savvidis}
P. G. Savvidis \textit{et al.}, Phys. Rev. Lett. \textbf{84} 1547 (2000)

\bibitem{huang}
R.\ Huang \textit{et al.}, Phys. Rev. B \textbf{61} R7854 (2000)

\bibitem{houdre}
R. Houdr\'{e} \textit{et al.}, Phys. Rev. Lett. \textbf{85} 2793 (2000); the structure of sample 1
is as described in this reference, except that it has only one quantum well.

\bibitem{stevenson}
R.M.\ Stevenson \textit{et al.}, Phys. Rev. Lett. \textbf{85} 3680 (2000)

\bibitem{immamoglu}
A. Immamoglu and R.J. Ram, Phys. Lett. A \textbf{214} 193 (1996)

\bibitem{tassone}
F. Tassone and Y.\ Yamamoto, Phys. Rev. B \textbf{59} 10830 (1999)

\bibitem{kira}
M. Kira \textit{et al.}, Phys. Rev. Lett. \textbf{79} 5170 (1997)

\bibitem{ciuti}
C. Ciuti \textit{et al.}, Phys. Rev. B \textbf{62} R4825 (2000)

\bibitem{kuwata}
M.\ Kuwata-Gonokami \textit{et al.}, Phys. Rev. Lett. \textbf{79} 1341 (1997)

\bibitem{quochi}
F.\ Quochi \textit{et al.}, Phys. Rev. B \textbf{159} R15594 (1999)

\bibitem{yariv}
A.\ Yariv, \textit{Quantum electronics} (John Wiley, New-York, 3rd edition, 1989), chap. 17

\bibitem{optamp}
Yi\ Mu and C.M.\ Savage, J. Opt. Soc. Am. B \textbf{9} 65 (1992); M.O. Scully and M.S. Zubairy,
\textit{Quantum Optics} (Cambridge University Press, 1995), chap. 21

\bibitem{nmc}
J.\ Rarity and C.\ Weisbuch, \textit{Microcavities and Photonic Bandgaps} (Kluwer, Dordrecht,1996),
p 43-57

\bibitem{yamamoto}
Y.\ Yamamoto and T.\ Kimura, IEEE J. Quant. Electron, QE-17 (1981); H van de Stadt, Astron.
Astrophys. \textbf{36} 341 (1974)

\bibitem{yuen}
H.P.\ Yuen and V.W.S. Chan, Opt. Lett. \textbf{8} 177 (1983)

\bibitem{stanley}
R.P. Stanley \textit{et al.}, Phys. Rev. B \textbf{55} R4867 (1997)

\bibitem{walls}
D.F. Walls and G.J. Milburn, \textit{Quantum optics} (Springer 1994), chap. 5

\bibitem{messin}
G. Messin \textit{et al.}, J.\ Phys.: Condens. Matter \textbf{11} 6069 (1999); H.\ Eleuch
\textit{et al.}, J.\ Opt. B: Quantum Semiclass. Opt. \textbf{1} 1 (1999)

\bibitem{note}
A correction is included to account for the non negligible value of the reflected field compared to
the local oscillator.

\end{thebibliography}
\end{document}